\documentclass[12pt] {article}
\usepackage{amsmath,epsfig}

\begin{document}
\pagestyle{myheadings} \markright{RFI mitigation}
\begin{center}
{\Large \bf RFI mitigation
with  phase-only adaptive beamforming}\\
\medskip
\vspace{5mm}
 P. A. Fridman\\
ASTRON,  Dwingeloo,P.O. Box 2, 7990 AA The Netherlands\\
\rm e-mail: fridman@astron.nl
\medskip
\vspace{5mm}

\end{center}

\begin{abstract}
  Connected radio interferometers are sometimes used in the tied-array mode: signals from antenna elements are coherently added and the sum signal applied to a VLBI backend or pulsar processing machine.  Usually there is no computer-controlled amplitude weighting in the existing radio interferometer facilities. Radio frequency interference (RFI) mitigation with  phase-only adaptive beamforming is proposed for this mode of observation. Small phase perturbations  are introduced  in each of the antenna's  signal. The values of these perturbations are optimized in such a way that the signal from  a radio source of interest is preserved and RFI signals  suppressed. An evolutionary programming algorithm is used for this  task. Computer simulations, made for both one-dimensional and two-dimensional array set-ups, show  considerable  suppression  of RFI and acceptable changes to the main array beam in the radio source direction.

\end{abstract}

\section{Introduction}
 Suppression of radio frequency interference (RFI) with adaptive  beamforming is widely used in radio astronomy, radar and telecommunications. The main idea behind many  algorithms proposed for use in radio astronomy consists of weighting the outputs of  array elements in  such a way as to create zero values in the synthesized array pattern in the direction of RFI and to keep the signal of interest (SOI), the radio source to be observed, in the maximum of the main lobe without significant loss of gain \cite{widrow1985, gab1992, krim1996}. During recent years there has been a growing interest in radio astronomy  for applying these methods of RFI mitigation   both to existing radio telescopes and to future generation projects. There are several specific features of the large  connected radio interferometers (RI) used in radio astronomy  such as  Westerbork Synthesis Radio Telescope (WSRT), Very Large Array (VLA) and Giant Metrewave Radio Telescope (GMRT)  which make the straightforward application of this adaptive beam-forming different and difficult when compared to  classic phased arrays:

1.  Connected RI are   highly sparse arrays.

2. Their main mode of operation is correlation processing.

3. Direction of arrival (DOA) of a signal of interest  is a known and time-dependent vector.

4. There is no computer-controlled amplitude weighting in the existing RI backend hardware.

5.  There is an auxiliary  {\it tied-array} facility which is used during VLBI and pulsar observations. The mode of observation is similar to that of standard phase arrays: the signals from the antennas are added but without amplitude weighting, because the antennas of RI are identical.
There is a phase-only control allowing coherent adding.  The RI works as a ``single dish".

6.  Noise-like  radio source signals  are usually  much weaker than  system noise (antenna + receiver) and RFI.

Phase-only adaptive nulling  is proposed for RFI mitigation during tied-array observations. Small phase perturbations  are introduced  to the   signals of every antenna. The values of these perturbations are optimized in such a way that the signal from  SOI is preserved and the RFI signals suppressed. This techniques has been widely discussed \cite{thompson1976, leavitt1976, stey1983, guisto1983, haupt1997, davis1998, smith1999} and is  well suited to  tied-array observations.
    \section{Narrow-band model of SOI and RFI}

There are two approaches to adaptive beam-forming: narrow-band (complex weighting of  amplitudes and phases) and wide-band (digital filtering, delay-tap  weighting). The narrow-band approach will be used in the following text.

Let us consider an equidistant M-element linear array. The $M$-dimensional array output vector $X(\theta )$, as a function of an angle, i. e., the complex amplitude of the temporal signal $x(t)=X(\theta )e^{j2\pi f_{0}t},$ consists of the following components:
     \begin{equation}
 X(\theta )=S(\theta _{0})+\sum_{n=1}^{N}RFI_{n}(\theta _{n})+N_{sys}
 \end{equation}

where $S(\theta _{0})$ is the signal vector corresponding to \ the plane wave coming from the direction $\theta _{0},RFI_{n}(\theta _{n})$ is the $n$th RFI vector, coming from any direction $\theta _{n},N_{sys}$ is the system noise vector. These three components are uncorrelated. Vector $S(\theta _{0})$ depends on the incidence angle $\theta _{0}$ of the plane wave, measured with respect to the normal to linear array
 \begin{equation}
S(\theta _{0})=[1,e^{-i\varphi _{0}},...e^{-i(M-1)\varphi _{0}}]^{T}
     \end{equation}
where phase shift $\varphi _{0}=(2\pi d/\lambda )\sin (\theta _{0}),d$ is the spacing between array elements, $\lambda $ is the wavelength. The phase of the first antenna is chosen to be equal to 0.

The beamformer, in general, consists of the complex weights $w_{m}e^{i\phi _{m}},m=1...M,$ which form the beamformer vector $W$
\begin{equation}
W=[1,w_{2}e^{i\phi _{2}},...w_{M}e^{i\phi _{M}}]^{T}.
    \end{equation}
The output of the phased array is
\begin{equation}
Y=W^{H}X.
    \end{equation}
The beamformer should satisfy both following requirements:

a) {\it steering capability}: the SOI is protected ($W^{H}S=g)$,  for a prescribed direction $\theta _{0},$ the response of the array is constant regardless of what values are assigned to the weights $W$;

b) the effects of RFI should be minimized.

The {\it minimum-variance distortionless response }(MVDR) beamforming algorithm, subject to this constraint when $g=1,$ is proposed in order to minimize the variance of the beamformer output\cite{capon1969} . The solution for $W$ in this case is
\begin{equation}
W_{MDVR}=R^{-1}S\left( \theta _{0}\right) [S\left( \theta _{0}\right) ^{H}R^{-1}S\left( \theta _{0}\right) ]^{-1},
    \end{equation}
where $R$ is the correlation matrix of $X.$

As was mentioned in the introduction, WSRT and other large radio astronomy arrays cannot use this algorithm in real time with the existing equipment because there are no amplitude control facilities. That is to say that  ``RFI nulling'' is limited to {\it  phase-only} control.

\section{Phase-only adaptive nulling}

Phase-only weights can be found to be the  solution to the following system of nonlinear equations:

\begin{eqnarray}
  Real\{\sum_{m=1}^{M}e^{i\phi _{m}}S_{m} =M\}  \\
  Imag\{\sum_{m=1}^{M}e^{i\phi _{m}}S_{m} =0\}   \\
  Real\{\sum_{m=1}^{M}e^{i\phi _{m}}RFI_{m} =0\} \\
  Imag\{\sum_{m=1}^{M}e^{i\phi _{m}}RFI_{m} =0\}.
\end{eqnarray}

$e^{i\phi _{m}}$ are the weights in our phase-only case. The vector $S$ is known and  is determined by the SOI coordinates. The construction of the  $RFI$ vector requires the knowledge of RFI's DOA, which may be known beforehand or could be obtained from \ the observed correlation matrix \ $\widehat{R_{ij}}=<x_{i}x_{j}>,i,j=1...M$, because SOI is always much weaker than RFI. But it is necessary to have a special correlator for this purpose in order to follow all rapid scintillations of  RFI which are usually averaged by the main radio interferometer correlator. So, in principle, the system of equations (6-9) can be solved and the phase corrections $\phi _{m}$  introduced into the phase control system. The optimal solution (5) can be used as a zero approximation for $\phi _{m}.$
The difficulties in implementing the solution of the system (6 - 9) are not mentioned here.

\section{Total power detector output}

A more practical method of calculating the phase corrections $\phi _{m}$ in the tied array case is proposed here. The tied array total power detector  output ($TPD_{TA}$) is
\begin{equation}
TPD_{TA}=<\int_{0}^{T}[x(t)]^{2}dt>=TPD_{sig}+TPD_{RFI}+TPD_{N},
    \end{equation}

where the total power components $TPD_{sig},TPD_{RFI},TPD_{N}$ correspond to the SOI, RFI and system noise, respectively, $<...>$ means the statistical expectation. The mean value of the signal component is
\begin{equation}
TPD_{sig}(\phi _{m})=<\int_{0}^{T}\left\{ \sum_{m=1}^{M}\cos [2\pi f_{0}t+2\pi (m-1)d\sin (\theta _{0})/\lambda +\phi _{m}]dt\right\}^{2} >,
    \end{equation}
the mean value of the n-th RFI component is
\begin{equation}
TPD_{RFI,n}(\phi _{m})=<\int_{0}^{T}\left\{ \sum_{m=1}^{M}A_{RFI,n}\cos [2\pi f_{0}t+2\pi (m-1)d\sin (\theta _{RFI,n})/\lambda +\phi _{m}]dt\right\}^{2} >,
    \end{equation}
the mean value of the system noise component  is constant. We assume also that $TPD_{sig}\ll TPD_{RFI,n}$  and the different $RFI_{n}$ are uncorrelated.

Considering the TPD output as a function of $M$ variables $\phi _{m},$ the following criterium for a ``good'' vector $\Phi =[\phi _{1}....\phi _{M}]^{T}$ can be proposed:
\begin{equation}
C(\Phi)=\frac{TPD_{sig}(\Phi )}{\sum_{n=1}^{N}TPD_{RFI,n}(\Phi )+TPD_{N}}\rightarrow \max .
    \end{equation}
The denominator $\sum_{n=1}^{N}TPD_{RFI,n}(\Phi )+TPD_{N}$ is the total TPD output under the assumption $TPD_{sig}<<TPD_{RFI,n}.$ The numerator $TPD_{sig}(\Phi)$ can be calculated for  each given $\Phi$ and $\theta _{0}$ (the DOA of SOI). Therefore, maximizing $C(\Phi)$ with a proper choice of $\Phi$, a higher signal-to-RFI-plus-noise ratio at the tied array output can be achieved.

This is a classic $M$-variable optimization problem which is difficult to solve by the common gradient methods because of the multimodality of $C({\bf \Phi }):$ there are many local (secondary) maximums and a searching algorithm will ``get stuck'' at  one of them without finding the global  maximum.

Genetic algorithms (GA) search the solution for the set of variables through the use of simulated evolution, i.e., {\it survival of the fittest} strategy. In contrast to the gradient algorithms which are, in general, calculus-based algorithms, GA, first introduced by \cite{holland1975}, exploits a {\it guided random techniques }during optimization procedure \cite{gold1989, mich1992, haupt1995}. The multimodality problem is successfully overcome by this algorithm.

A simplified block diagram of  GA implementation in a radio interferometer is depicted in Figure 1. A phase control subsystem introduces a certain initial phase distribution $\Phi _{0}$ corresponding to  radio source  coordinates and preliminary phase calibration corrections.  The output of the TPD is then continiously measured and used to supply the GA program with the data (cost function samples) which monitor the performance of the tied array with respect to RFI. The GA uses these data to calculate new phases $\Phi _{m}$ with the aim to maximize $C(\Phi).$ These new phases are introduced into the phase control subsystem after each iteration and a new value of the TPD output signal is used for the next step. Thus the feedback loop, {\it phase control subsystem - TPD - GA}, maintains the low value of  $TPD_{TA}(\Phi)$ and therefore the high value of the fitness function $C(\Phi)$, i. e.,  the high signal-to-RFI ratio.

    \section{Computer simulation }
Computer simulation was performed to illustrate the effectiveness of the phase-only nulling in  RFI mitigation.

First, a 14-element half-wavelength linear array was modelled. The SOI direction is equal to $0^{\circ },$ and there are two RFI signals: one at the angle $-20.1^{\circ }$ , and the other at the angle $+10.015^{\circ }.$ Figure 2 shows in logarithmic scale  the quiescent (dash line) and adapted (solid line) array patterns. The significant  suppression of RFI with the adapted pattern is clearly visible, while the quiescent pattern has the secondary lobe maximums at the RFI positions. Figure 3 shows the corresponding array phase  distribution.

Sparse 14-element array  patterns are represented in Figures 4, 5, 6 and 7. The distance between the elements is $144m$ and the central frequency is $1420MHz$. The lobes are much narrower than for the half-wavelength array, so the  different figures are given to illustrate the result of phase-only nulling. Figure 4 (a linear scale presentation of the pattern) is given here to illustrate the loss and distortions of the main lobe. This is more visible in the linear scale, whereas  RFI suppression is better seen in the logarithmic scale. Figure. 5 - the quiescent and adapted patterns around angle $0^{\circ },$ logarithmic scale, Figures   6, 7 - the  same patterns around the directions of RFI1 (DOA$=+10.015^{\circ })$ and RFI2 (DOA$=-20.1^{\circ }).$ The corresponding array phase distribution is shown in Fig. 8.

The subsequent figures illustrate this phase-only nulling  for a two-dimensional planar array.

Adaptive nulling was simulated for the  half-wavelength array with 10x10 elements, the central frequency is equal to $1420MHz$. Rectangular coordinates $a1$ and $a2$ are angles measured from the $x$ and $y$ axes, respectively, to the line from the array to the radio source; thus   the SOI is at the  zenith, with coordinates  $(90^{\circ }, 90^{\circ }).$ Coordinates of  RFI were chosen so as to put them on the maximums of the secondary lobes, the values of RFI suppression are shown in the captions. The following sequence of figures is given:

Figure 9: normalized (A($90^{\circ },90^{\circ })$=1) quiescent array with indicated RFI positions;

Figure 10:  array's pattern after adaptation;

Figure 11:  this array's phase distribution after adaptation.

\section{Conclusions}

1. Existing large radio interferometers (WSRT, VLA, GMRT) have only phase control facilities and the {\it real-time} adaptive nulling in the RFI direction should take this constraint into account .

2. The  total power detector at the tied-array output can be used for  phase-only RFI mitigation as an indicator of the level of RFI.

3. The Genetic Algorithm is a convenient tool for  cost function maximization during the search for the optimal array phase distribution.

4. Computer simulations show  significant RFI mitigation for the sparse linear array in the narrow-band  approximation ($\Delta f/f_{0}<<1$).

5. Phase-only nulling can also be used for  real-time RFI mitigation at the station's level in   new projects such as ATA, LOFAR and SKA.

\newpage
\begin{figure}
 \includegraphics[height=9.0cm,width=12.0cm]{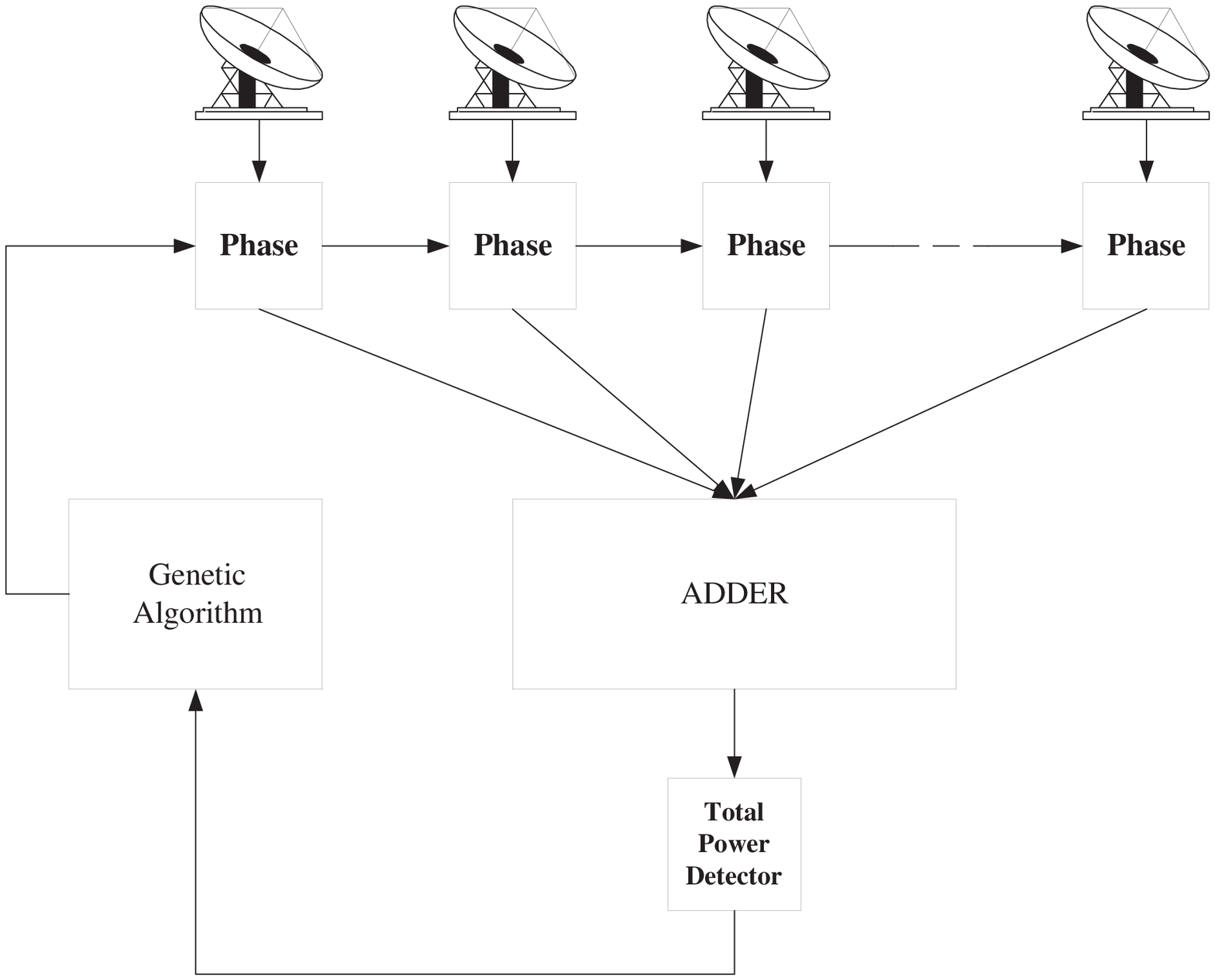}
 \caption{Block diagram of TPD adaptive phase control}
\end{figure}

\begin{figure}
 \includegraphics[height=9.0cm,width=12.0cm]{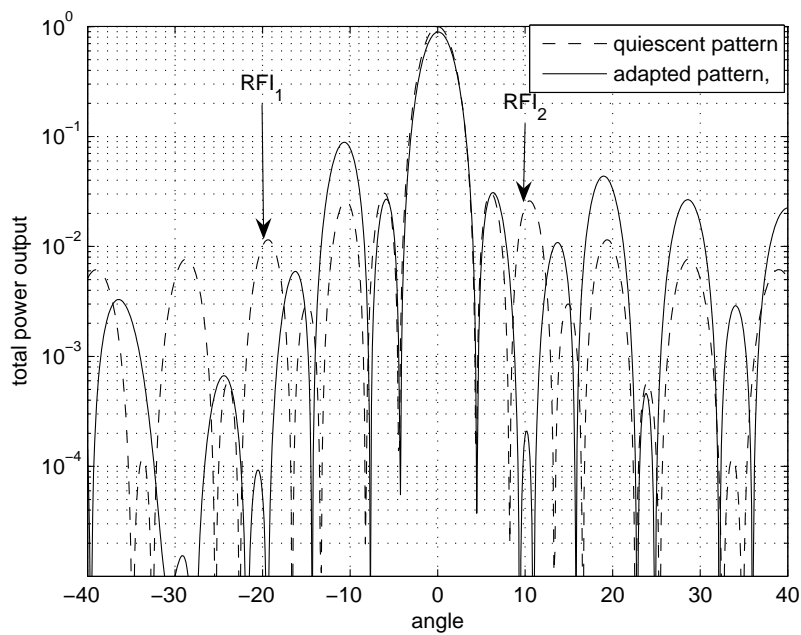}
 \caption{Quiescent (dash line) and adapted (solid line) 14-element half-wavelength array pattern, logarithmic scale; two nulls in the adapted pattern at $-20.1^{\circ}$ and $+10.015^{\circ}$.}

\end{figure}

\begin{figure}
 \includegraphics[height=9.0cm,width=12.0cm]{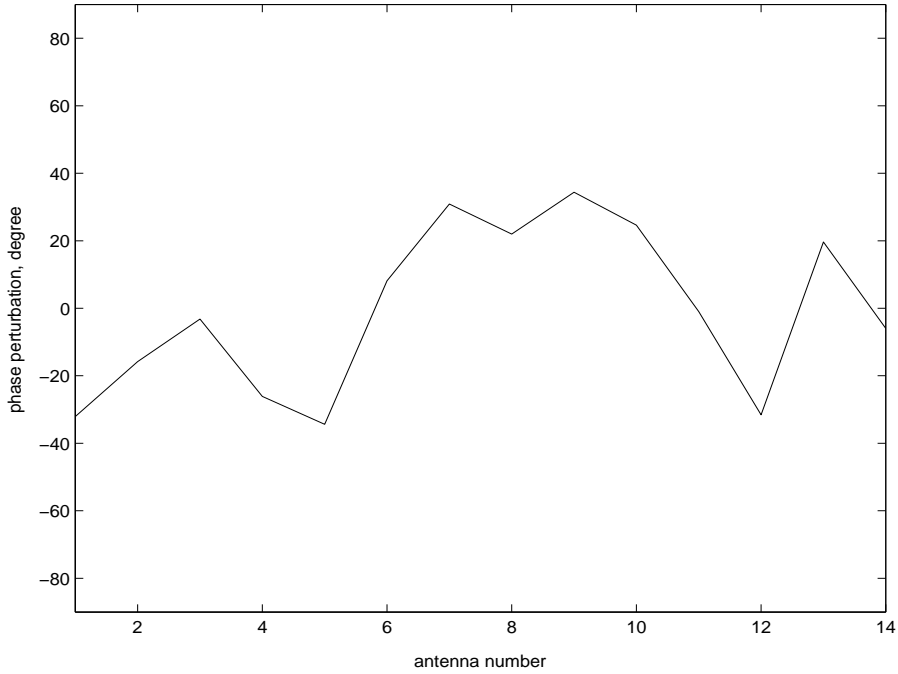}
\caption{Phase perturbations corresponding to the  adapted array pattern in Figure 2.}
\end{figure}

\begin{figure}
 \includegraphics[height=9.0cm,width=12.0cm]{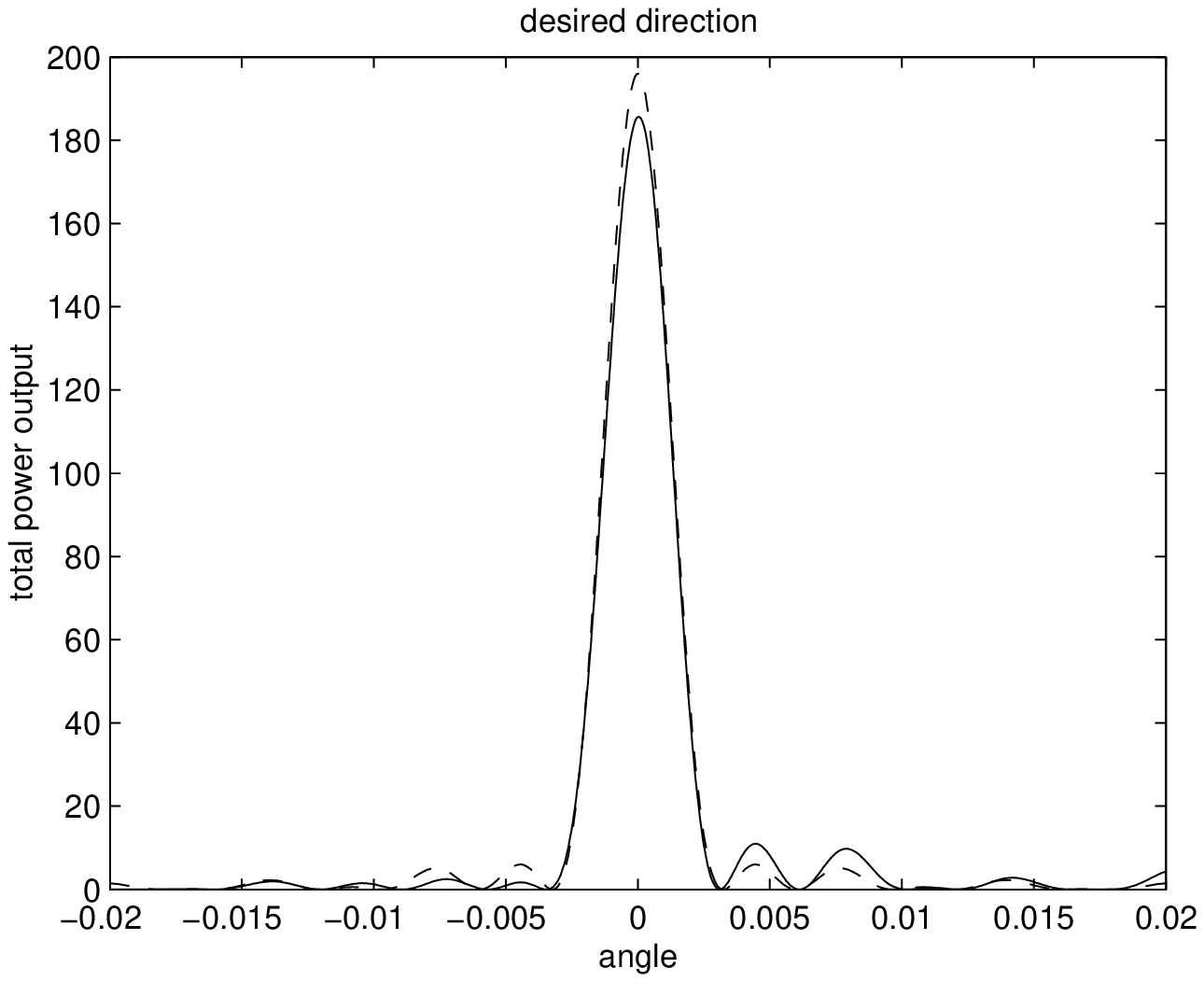}
 \caption{Quiescent (dash line) and adapted (solid line) 14-element array pattern, spacing=144m, central frequency=1420MHz,  main beam, linear scale.}
\end{figure}

\begin{figure}
 \includegraphics[height=9.0cm,width=12.0cm]{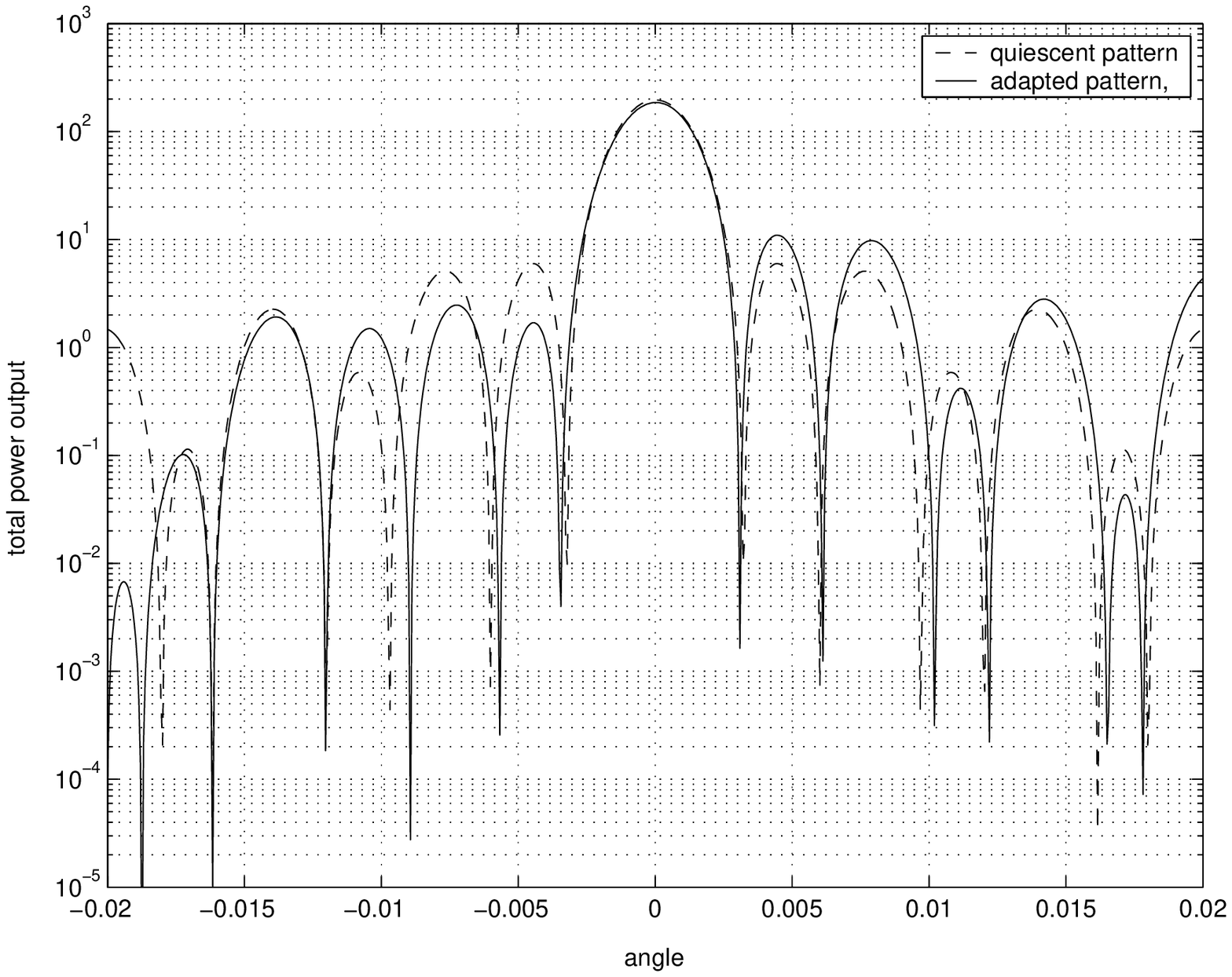}
 \caption{Quiescent (dash line) and adapted (solid line) 14-element array pattern, spacing=144m, central frequency=1420MHz, main beam, logarithmic scale.}
\end{figure}

\begin{figure}
 \includegraphics[height=9.0cm,width=12.0cm]{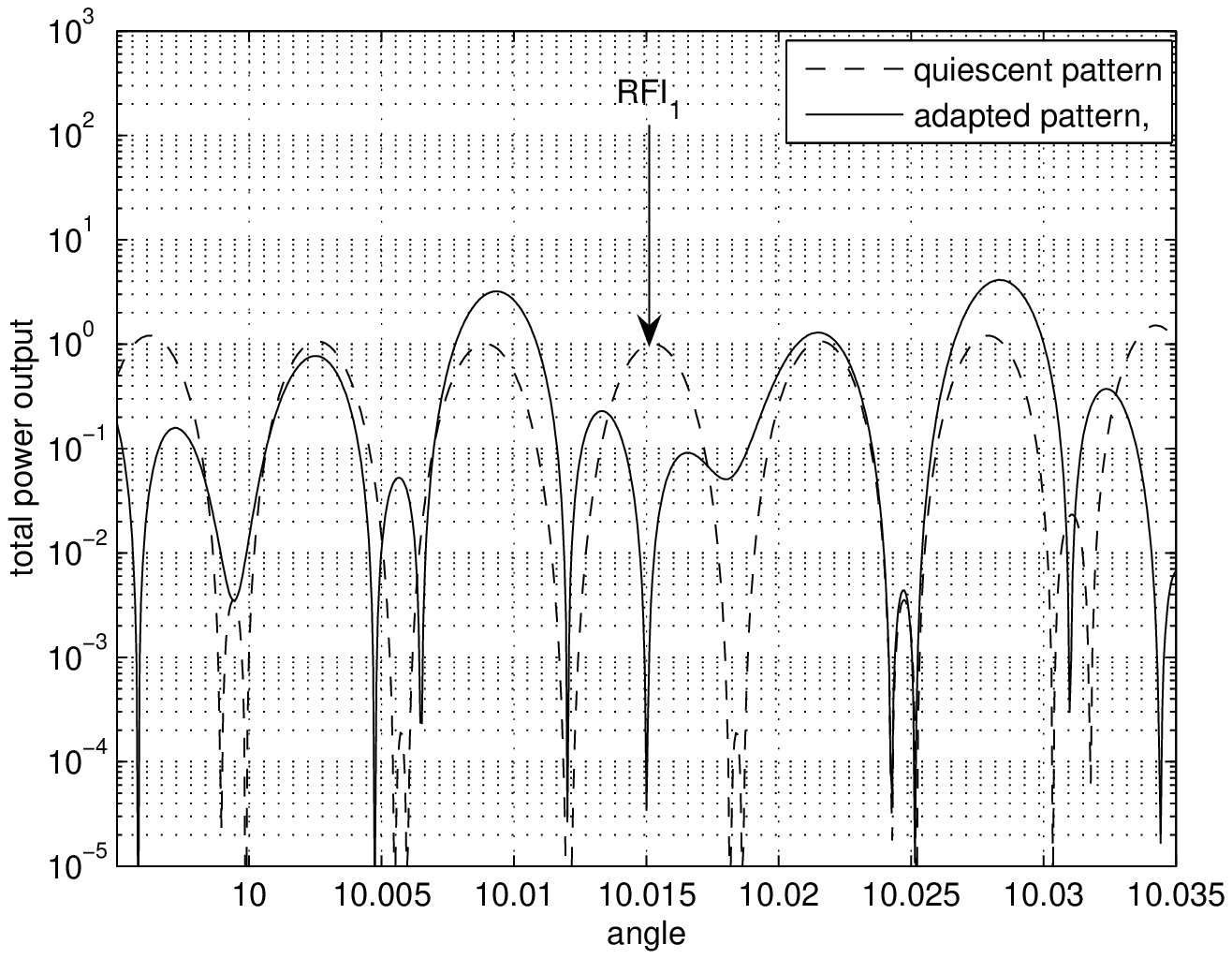}
 \caption{Quiescent (dash line) and adapted (solid line) 14-element array pattern, spacing=144m, central frequency=1420MHz, direction $+10.015^{\circ}$, logarithmic scale.}
\end{figure}

\begin{figure}
 \includegraphics[height=9.0cm,width=12.0cm]{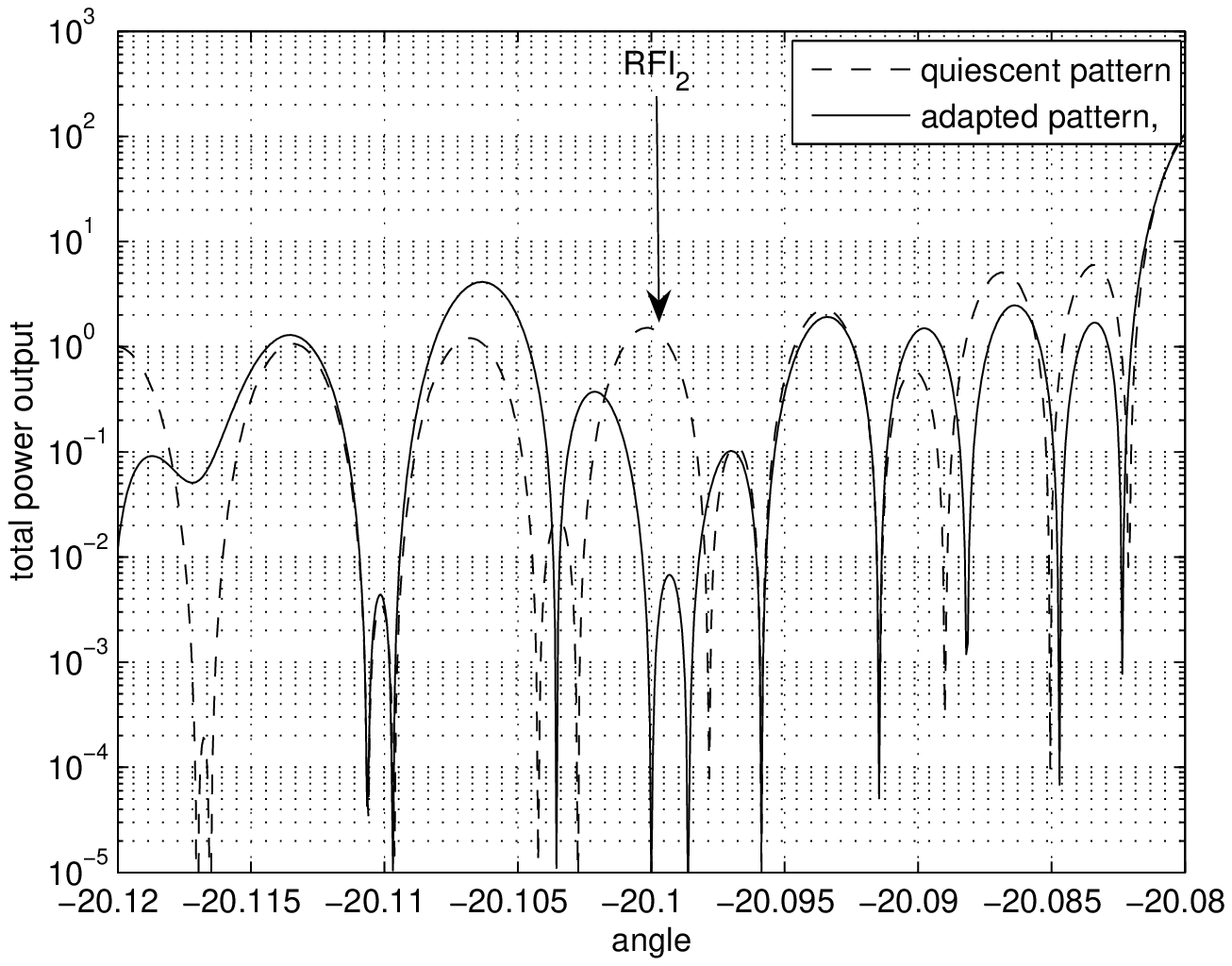}
 \caption{Quiescent (dash line) and adapted (solid line) 14-element array pattern, spacing=144m, central frequency=1420MHz, direction $-20.1^{\circ}$, logarithmic scale.}
\end{figure}

\begin{figure}
 \includegraphics[height=9.0cm,width=12.0cm]{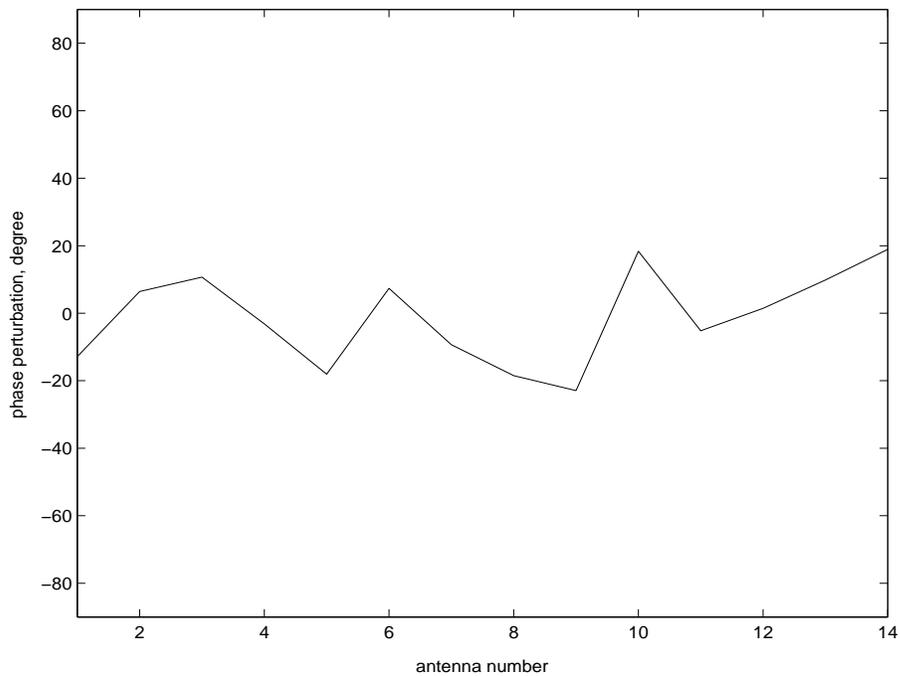}
 \caption{Phase perturbations corresponding to the adapted array pattern in Figures 5, 6 and 7.}
\end{figure}

\begin{figure}
 \includegraphics[width=15cm,height=9cm]{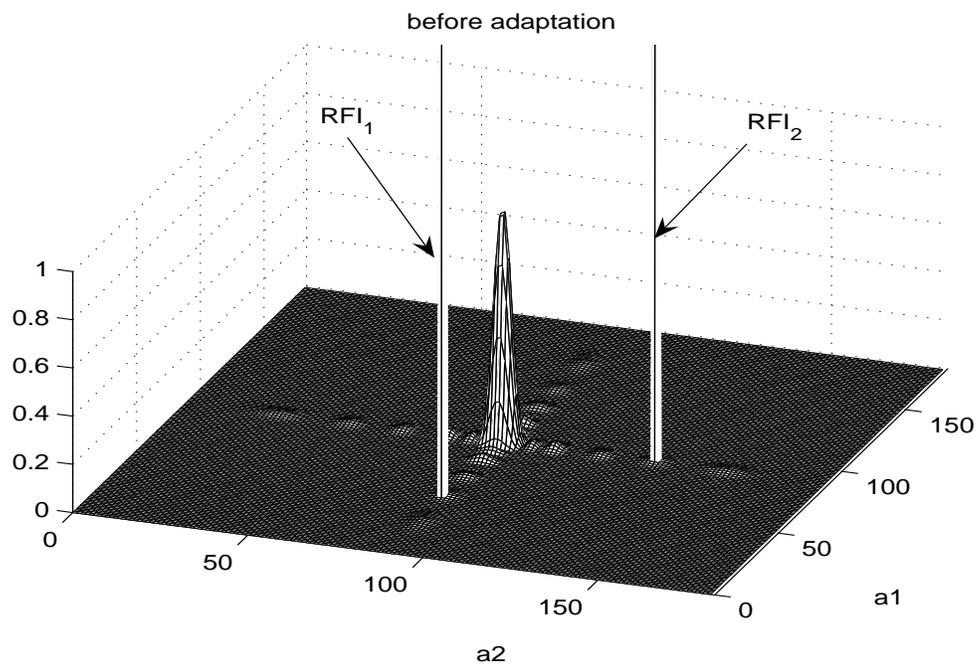}
 \caption{Two-dimensional 10x10-element half-wavelength, {\it quiescent} array pattern, linear scale, central frequency=1420MHz, RFI-1 at $[45^{\circ},90^{\circ}]$,  RFI-2 at $[90^{\circ},135^{\circ}]$, directions of RFI  coincide with the maximums of the sidelobes.}
\end{figure}
\newpage
\begin{figure}
 \includegraphics[width=15cm,height=9cm]{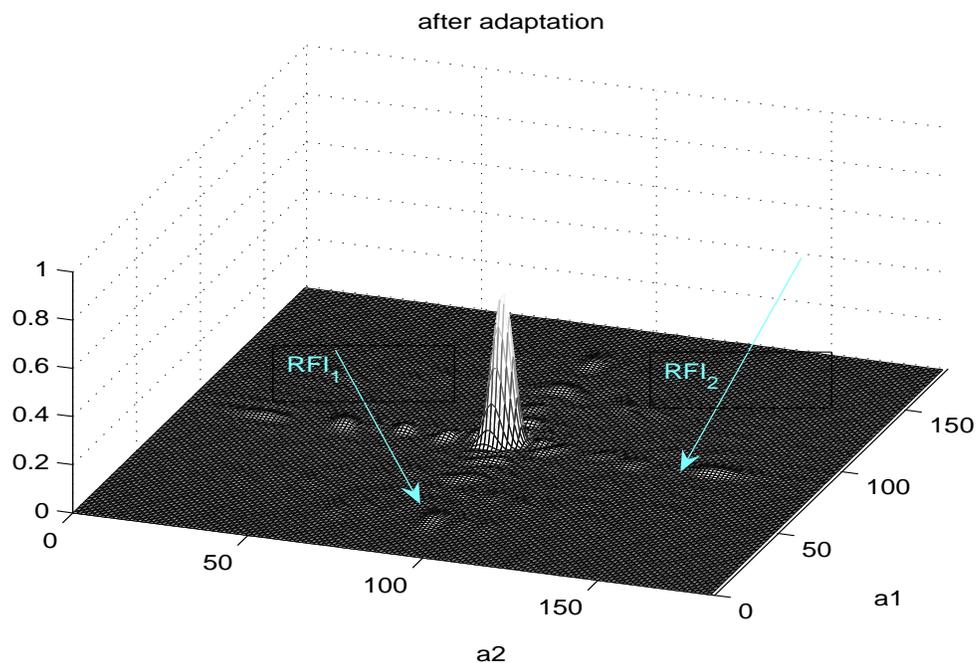}
 \caption{Two-dimensional 10x10-element half-wavelength, {\it adapted} array pattern, central frequency=1420MHz, RFI-1 at $[45^{\circ},90^{\circ}]$,  RFI-2 at $[90^{\circ},135^{\circ}]$, linear scale; RFI-1 suppression=106.2dB, RFI-2 suppression=103.1dB.}
\end{figure}

\begin{figure}
 \includegraphics[width=15cm,height=12cm]{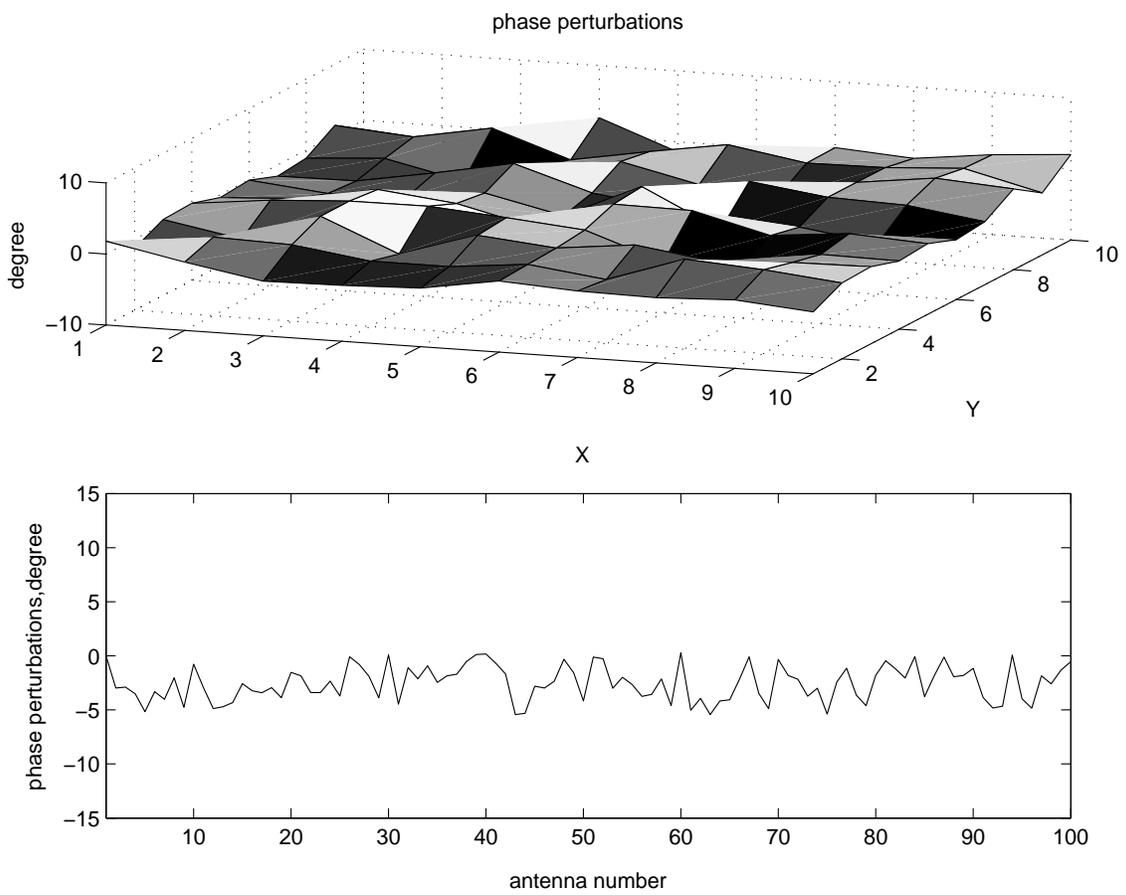}
 \caption{Phase perturbations corresponding to the adapted array  pattern in Fig. 10: upper panel shows a 3D-presentation of the phase surface, lower panel shows the phase distribution in linear order.}
\end{figure}
\end{document}